\documentclass{pasj00}

\draft

\begin{document}
\SetRunningHead{S. Kato}{Disk Oscillations in Vertically Terminated Disks}
\Received{2010/00/00}
\Accepted{2010/00/00}

\title{Trapped, Two-Armed, Nearly Vertical Oscillations in Disks with 
Toroidal Magnetic Fields II: Effects of Finite Thickness}

\author{Shoji \textsc{Kato}}
\affil{2-2-2 Shikanoda-Nishi, Ikoma-shi, Nara, 630-0114}
\email{kato@gmail.com, kato@kusastro.kyoto-u.ac.jp}

%


\maketitle

\begin{abstract}

We examine radial trapping of two-armed ($m=2$) vertical p-mode oscillations in geometrically 
thin relativistic disks which are vertically isothermal 
but terminated at a certain height by the presence of hot and low-density corona.
The disks are assumed to be subject to toroidal magnetic fields.
The oscillations are classified by $n$, a number related to the node number of oscillations
in the vertical direction and starting from $n=1$.
In modes with $n=1$, the frequencies of trapped oscillations 
depend little on the height of termination, but in modes with $n=2,3,...$ the frequencies 
decrease and the radial extends of trapped region become wide, 
as the termination height decreases.
This study is a preparation to examine whether these oscillations can describe kilo-hertz
quasi-periodic oscillations (kHz QPOs),
horizontal branch oscillation (HBOs), and their correlations.
    
\end{abstract}

\section{Introduction}

Kilo-hertz quasi-periodic oscillations (kHz QPOs) observed in neutron-star low-mass
X-ray binaries (NS LMXBs) are one of promising tools to investigate the innermost structure of
relativistic disks and to evaluate the mass and spin of the central neutron stars.
Although there is no general consensus on the origin of the kHz QPOs, one of promising origins is 
disk oscillations trapped in the innermost region of relativistic disks.

There are many disk oscillations which are trapped in the innermost relativistic
disks (for review, see Wagoner 1999, Kato 2001, Kato et al. 2008).
Recently, Kato (2010) showed that in addition the trapped oscillations reviewed in the above review works,
two-armed $(m=2)$ vertical p-mode oscillations can also be trapped in the innermost region of disks
with frequencies relevant to kHz QPOs.
The characteristics of these two-armed oscillations were examined by Kato (2011a, paper I)
in the case where the disks are subject to toroidal magnetic fields.
Subsequently, Kato (2011b) showed that these oscillations can naturally describe the
correlated frequency change of the twin kHz QPOs observed in neutron-star LMXBs.

In the above model of twin kHz QPOs, the frequency change of the QPOs are assumed to be due to
time changes of acoustic and Alfv\'{e}n speeds in disks.
In addition to them, however, one more possible causes of frequency change will be conceivable.
This is time change of vertical disk thickness.
The time change of vertical disk thickness will be expected, since geometrically thin disks in 
LMXBs will be surrounded by hot corona and its transition height will vary with time by a change
of evaporation efficiency by change of disk state.
In this sense, it will be of interest to examine how the change of transition height 
affects the frequencies of trapped two-armed vertical p-mode oscillations.

Based on the above considerations, we examine in this paper the characteristics of 
two-armed ($m=2$) vertical p-mode oscillations in the case where the disks are vertically isothermal but
terminated at a certain height by presence of hot low-density corona.
Application of the results to observed frequency correlations of QPOs will be made in a subsequent
paper (Kato 2012).
   
\section{Unperturbed Disks and Equations Describing Disk Oscillations}

The unperturbed disks are isothermal in the vertical direction.
We assume that the disks are terminated at a certain height by the presence of a hot corona.
When we consider oscillations, a revelant boundary condition is imposed at the height of transition
between the disk and the corona, as will be discussed later.
Except for this, the disks considered in this paper are the same as those considered by paper I.
That is, the disks are geometrically thin and relativistic.
For mathematical simplicity, however, the effects of general relativity are taken into 
account only when we consider radial distributions of $\Omega(r)$, $\kappa(r)$, and $\Omega_\bot(r)$, 
which are, in turn, the angular velocity of disk rotation, the epicyclic frequencies 
in the radial and vertical directions. 
Except for them, the Newtonial formulations are adopted.
Since geometrically thin disks are considered, $\Omega$, is approximated to be the relativistic 
Keplerian angular velocity, $\Omega_{\rm K}(r)$, when its numerical values are necessary.
Here, $r$ is the radial coordinate of cylindrical ones ($r$,$\varphi$,$z$), where the $z$-axis is
perpendicular to the disk plane and its origin is the disk center.
Functional forms of $\Omega_{\rm K}(r)$, $\kappa(r)$, and $\Omega_\bot(r)$ are given 
in many literatures (e.g., Kato et al. 2008).

\subsection{Unperturbed Disks with Toroidal Magnetic Fields}

The equilibrium disks are axisymmetric and vertically isothermal with toroidal magnetic fields.
The fields are assumed to be purely toroidal:
\begin{equation}
      \mbox{\boldmath $B$}_0(r,z)=[0,B_0(r,z),0],
\label{2.1}
\end{equation}
and distributed in such a way that the Alfv\'{e}n speed, $c_{\rm A}$, 
is constant in the vertical direction, i.e., $(B_0^2/4\pi\rho_0)^{1/2}=$ const. in the vertical
direction, where $\rho_0(r,z)$ is the density in the unperturbed disks.

Since both the isothermal acoustic speed, $c_{\rm s}$, and the Alfv\'{e}n speed, $c_{\rm A}$, are 
constant in the vertical direction, the integration of the hydrostatic balance in the vertical
direction gives (e.g., Kato et al. 1998)
\begin{equation}
    \rho_0(r,z)=\rho_{00}(r){\rm exp}\biggr(-\frac{z^2}{2H^2}\biggr), \quad{\rm and}\quad
    B_0(r,z)=B_{00}(r){\rm exp}\biggr(-\frac{z^2}{4H^2}\biggr),
\label{2.3}
\end{equation}
where the scale height $H$ is related to $c_{\rm s}$, $c_{\rm A}$, and $\Omega_\bot$ by
\begin{equation}
      H^2(r)=\frac{c_{\rm s}^2+c_{\rm A}^2/2}{\Omega_\bot^2}.
\label{2.4}
\end{equation}
The ratio $c_{\rm A}^2/c_{\rm s}^2$ is a parameter describing the disk structure.
It is a function of the radius in general, 
but in this paper it is taken to be constant throughout the trapped region of oscillations.
This is because the trapped region is found to be narrow except for some cases where 
$c_{\rm A}^2/c_{\rm s}^2$ is large (see figure 7).

The disks described above are assumed to be terminated at a certain height, say $z_{\rm s}$, 
by a high-temperature and low-density corona.
The ratio $z_{\rm s}/H$ is one of parameters describing the disk, which is also assumed to be constant 
in the trapped region.

\subsection{Equations Describing Disk Oscillations}

Small amplitude perturbations are superposed on the equilibrium disk described above.
The velocity perturbation over rotation is denoted by ($u_r$, $u_\varphi$, $u_z$), and
the perturbed part of magnetic field over the unperturbed one by ($b_r$, $b_\varphi$, $b_z$).

The azimuthal and time dependences of the perturbed quantities are taken to be proportional to 
exp$[i(\omega t-m\varphi)]$, where $\omega$ and $m$ are the frequency and the azimuthal wavenumber 
of perturbations, respectively.
In this paper we are interested only in two-armed ($m=2$) oscillations, but 
$m$ is retained here and hereafter without specifying so that we can trace back the terms coming from 
$m$ in the final results. 
The perturbations are assumed to be local in the sense that their characteristic radial scale, 
$\lambda$, is shorter than the characteristic radial scale of the disks, $\lambda_{\rm D}$.
The latter is of the order of $r$, i.e., 
i.e., $\lambda<\lambda_{\rm D}(\sim r)$.
By using this approximation, we neglect such quantities as
$d{\rm ln}\rho_{00}/d{\rm ln}r$, $d{\rm ln}B_{00}/d{\rm ln}r$, $d{\rm ln}H/d{\rm ln}r$,
and $d{\rm ln}\Omega/d{\rm ln}r$, compared with terms of the order of $r/\lambda$.
Then, the $r$-, $\varphi$-, and $z$-components of equation of motion are reduced to
(e.g., paper I)
\begin{equation}
   i(\omega-m\Omega)u_r-2\Omega u_\varphi=-\frac{\partial h_1}{\partial r}-
          c_{\rm A}^2\frac{\partial}{\partial r}\biggr(\frac{b_\varphi}{B_0}\biggr),
\label{fluidr1}
\end{equation}
\begin{equation}
    i(\omega-m\Omega)u_\varphi+\frac{\kappa^2}{2\Omega}u_r=0,
\label{fluidvarphi1}
\end{equation}
\begin{equation}
    i(\omega-m\Omega)u_z=-\biggr(\frac{\partial}{\partial z}+\frac{c_{\rm A}^2}{2c_{\rm s}^2}
        \frac{z}{H^2}\biggr)h_1
        -c_{\rm A}^2\biggr(\frac{\partial}{\partial z}-\frac{z}{H^2}\biggr)\biggr(\frac{b_\varphi}{B_0}\biggr)
        -i\frac{m}{r}c_{\rm A}^2\biggr(\frac{b_z}{B_0}\biggr).
\label{fluidz1}
\end{equation}
In the above equations, $h_1$ defined by $h_1=p_1/\rho_0=c_{\rm s}^2\rho_1/\rho_0$ has been
introduced by assuming adiabatic perturbations, where $p_1$ and $\rho_1$ are, respectively,
pressure and density perturbations over $p_0$ and $\rho_0$.

Similarly, the $r$-, $\varphi$-, and $z$-components of induction equation are reduced to
\begin{equation}
     i(\omega-m\Omega)\frac{b_r}{B_0}=-i\frac{m}{r}u_r,
\label{magr1}
\end{equation}
\begin{equation}
     i(\omega-m\Omega)\frac{b_\varphi}{B_0}=r\frac{d\Omega}{dr}\frac{b_r}{B_0}
            -\frac{\partial u_r}{\partial r}
            -\biggr(\frac{\partial}{\partial z}-\frac{z}{2H^2}\biggr)u_z,
\label{magvarphi1}
\end{equation}
\begin{equation}
     i(\omega-m\Omega)\frac{b_z}{B_0}=-i\frac{m}{r}u_z.
\label{magz1}
\end{equation}
Finally, the equation of continuity is reduced to
\begin{equation}
     i(\omega-m\Omega)h_1=-c_{\rm s}^2\biggr[\frac{\partial u_r}{\partial r}
            +\biggr(\frac{\partial}{\partial z}-\frac{z}{H^2}\biggr)u_z\biggr].
\label{continuity1}
\end{equation}

Now, we further simplify equations (\ref{fluidz1}) and (\ref{magvarphi1}).
The last term, $-i(m/r)c_{\rm A}^2(b_z/B_0)$, of equation (\ref{fluidz1}) 
can be expressed in terms of $u_z$ by using equation (\ref{magz1}).
The result shows that the term of $-i(m/r)c_{\rm A}^2(b_z/B_0)$ is smaller than 
the left-hand term, $i(\omega-m\Omega)u_z$, by a factor of $c_{\rm A}^2/r^2\Omega^2$.
Considering this, we neglect the last term on the right-hand side of 
equation (\ref{fluidz1}).
Next, we consider equation (\ref{magvarphi1}).
The first term on the right-hand side, $r(d\Omega/dr)(b_r/B_0)$, is 
smaller than the second term, $-\partial u_r/\partial r$, by
a factor of $\lambda/r$, which can be shown by expressing $b_r$ in terms of $u_r$ by
using equation (\ref{magr1}).
Hence, we neglect the term in the following analyses.

After introducing the above approximations into equations (\ref{fluidz1}) and 
(\ref{magvarphi1}), we multiply $i(\omega-m\Omega)$ to equation
(\ref{fluidz1}) in order to express $h_1$ and $b_\varphi/B_0$ in equation (\ref{fluidz1})
in terms of $u_z$ and $u_r$ by using equation (\ref{continuity1}) and (\ref{magvarphi1}).
Then, after changing independent variables from ($r$, $z$) to ($r$, $\eta$), where $\eta$
is defined by $\eta=z/H$, we have (paper I)
\begin{equation}
    \biggr[\frac{\partial^2}{\partial \eta^2}-\eta\frac{\partial}{\partial \eta}
       +\frac{(\omega-m\Omega)^2-\Omega^2_\bot}{c_{\rm s}^2+c_{\rm A}^2}H^2\biggr]u_z 
            +H\biggr[\frac{\partial}{\partial\eta}
            -\frac{c_{\rm A}^2/2}{c_{\rm s}^2+c_{\rm A}^2}\eta\biggr]\frac{\partial u_r}{\partial r}=0.
\label{wave}
\end{equation}
This is the basic wave equation to be solved in this paper and the same as that in paper I.

\section{Nearly Vertical Oscillations}

Equation (\ref{wave}) is now solved by approximately decomposing it into two
equations describing oscillatory behaviors in vertical and radial directions as done in paper I, following 
Silbergleit et al. (2001) and Ortega-Rodrigues (2008).

\subsection{Boundary Condition at Disk Surface and Vertical Eigen-functions}

As mentioned before, the oscillations which we are interested in here are nearly vertical in the 
lowest order of oscillations (i.e., vertical p-mode oscillations).
The main terms in equation (\ref{wave}) are thus those of the first brackets, and
the terms of the second brackets are small perturbed quantities (see paper I).
Although the terms of the second brackets are small quantities, they are of importance
to determine the wave trapping in the radial direction, as is shown in the next section.

First, we should notice that the quantity with $(\omega-m\Omega)^2-\Omega_\bot^2$ in the first
brackets depends weakly on radius $r$. 
Hence, in order to consider this weak $r$-dependence of the quantity 
as a small perturbed one, the third term in the first brackets of equation (\ref{wave}) is
now expressed as
\begin{equation}
       \frac{(\omega-m\Omega)^2-\Omega_\bot^2}{c_{\rm s}^2+c_{\rm A}^2}H^2
        =\biggr[\frac{(\omega-m\Omega)^2-\Omega_\bot^2}{c_{\rm s}^2+c_{\rm A}^2}H^2\biggr]_{\rm c}
           +\epsilon(r),
\label{eigenvalue0}
\end{equation}
where the subscript c represents the value at capture radius, $r_{\rm c}$, 
which is the outer boundary of the radial propagation region of oscillations and will be determined later.
By definition, $\epsilon$ vanishes at $r_{\rm c}$, i.e., $\epsilon(r_{\rm c})=0$.
The magnitude of $\epsilon(r)$ is found from equation (\ref{eigenvalue0}) when $r_{\rm c}$ and $\omega$ 
are determined later by solving an eigen-value problem in the radial direction 
(see subsequent sections, especially see the final paragraph of section 5 and figure 9). 

The fact that $r_{\rm c}$ is really the capture radius of oscillations can be found from
the following considerations.
The final results show that $\epsilon(r)$ is a small positive quantity which monotonically decreases outwards
in the region of $r<r_{\rm c}$ and vanishes at $r_{\rm c}$ (see figure 9), i.e.,
$\epsilon(r)> 0$ for $r<r_{\rm c}$.
This implies that $(\omega-m\Omega)^2-\Omega_\bot^2$ is positive throughout the region of $r<r_{\rm c}$,
when $[(\omega-m\Omega)^2-\Omega_\bot^2]_{\rm c}$ is positive. 
This further implies that $(\omega-m\Omega)^2-\kappa^2$ is also positive in the region since 
$\kappa^2$ is always smaller than $\Omega_\bot^2$.
The facts of $(\omega-m\Omega)^2-\kappa^2>0$ and $\epsilon>0$ in the region of $r<r_{\rm c}$
means that the region is a propagation region of oscillations and the oscillations are trapped there 
[see equation (\ref{wave-eq-radial})].

If the term of $\epsilon(r)$ is transported to terms of small perturbations, equation (\ref{wave})
in the lowest order of approximations is written in the form:
\begin{equation}
    \frac{\partial^2}{\partial \eta^2}u_z^{(0)}-\eta\frac{\partial}{\partial \eta}u_z^{(0)}
    +\biggr[\frac{(\omega-m\Omega)^2-\Omega^2_\bot}{c_{\rm s}^2+c_{\rm A}^2}H^2\biggr]_{\rm c}u_z^{(0)}
    = 0,
\label{zeroth}
\end{equation}
where the superscript (0) is attached to $u_z$ in order to emphasize that it is the quantity of 
the lowest order of approximations.
To solve equation (\ref{zeroth}), we must impose boundary conditions.
In the case of isothermal disks which extend infinitely in the vertical direction, 
we impose that $u_z^{(0)}$ does not grow exponentially at $z=\pm \infty$ (see Okazaki et al. 1987).
In the present case of finite thickness of disks, however, a natural boundary condition to be imposed 
at $z=z_s$ (half-thickness of the disks) will be vanishing of the Lagrangian perturbation of the total 
pressure, i.e.,
$\delta p_{\rm tot}=0$, where $p_{\rm tot}=p +B^2/8\pi$.
This is because outside the surface acoustic perturbations (more accuratly, fast mode of MHD perturbations) 
will be propagated away quickly by the presence of high-temperature and low-density corona.

The condition of $\delta p_{\rm tot}=0$ can be written for nearly vertical oscillations as
\begin{equation}
    p_1+\frac{B_0^2}{4\pi}\frac{b_\varphi}{B_0}+\xi_z\frac{\partial}{\partial z}
        \biggr(p_0+\frac{B_0^2}{8\pi}\biggr)=0,
\label{boundary1}
\end{equation}
where $\xi_z$ is the vertical component of displacement vector, $\mbox{\boldmath $\xi$}$,
associated with the perturbations.
Since $p_1=\rho_0 h_1$, $i(\omega-m\Omega)\xi_z=u_z$, and the vertical hydrostatic balance gives
$\partial(p_0+B_0^2/8\pi)/\partial z=-\rho_0\Omega_\bot^2z$, equation (\ref{boundary1}) can be written as 
\begin{equation}
       h_1+c_{\rm A}^2\biggr(\frac{b_\varphi}{B_0}\biggr)-\xi_z\Omega_\bot^2 z=0.
\label{boundary2}
\end{equation}
Since we are considering nearly vertical oscillations, in the lowest order of approximation, we have
$i(\omega-m\Omega)h_1=-c_{\rm s}^2(\partial/\partial z-z/H^2)u_z$ from equation (\ref{continuity1}) and 
$i(\omega-m\Omega)(b_\varphi/B_0)=-(\partial/\partial z-z/2H^2)u_z$ from equation (\ref{magvarphi1}).
Hence, substituting these relations into equation (\ref{boundary2}), we have,
with the help of equation (\ref{2.4}), 
\begin{equation}
    \frac{\partial u_z^{(0)}}{\partial \eta}=0 \quad 
      {\rm at} \quad \eta=\pm \eta_{\rm s} \biggr(\equiv \pm\frac{z_{\rm s}}{H}\biggr).
\label{boundary3}
\end{equation}

Let us introduce a symbol $K_{n,{\rm s}}$ defined by
\begin{equation}
     \biggr[\frac{(\omega-m\Omega)^2-\Omega_\bot^2}{c_{\rm s}^2+c_{\rm A}^2}H^2\biggr]_{\rm c}=K_{n, {\rm s}}.
\label{eigenvalue}
\end{equation}
Then, by solving equation (\ref{zeroth}) with the boundary condition (\ref{boundary3}),
we have a discrete set of $K_{n,{\rm s}}$ as eigen-values.
The value $K_{n,{\rm s}}$ depends on the node number of oscillations in the vertical direction
(characterized by subscript n) and the height of disk surface (characterized by subscript s).
In the lowest mode of oscillation with respect to node number in the vertical direction,
we have $u_z^{(0)}=$ const. and $K_{n,{\rm s}}=0$, independent of s.
In this paper as in paper I, this mode is labelled by $n=1$ not by $n=0$, i.e., $K_{1,{\rm s}}=0$, and
$u_{z,(1,s)}^{(0)}(={\rm const}.)$ has no node in the vertical direction.\footnote{
The reason why this fundamental mode is labelled  by $n=1$ in spite of $u_z$ having no node 
in the vertical direction is that mode classification is made in many cases 
by the node number of $u_r$, and $u_r$ has one more node(s) in the vertical direction, compared with
$u_z$.
}
Figure 1 shows the eigen-value, $K_{n,{\rm s}}$, of three modes of $n=1$, 2, and 3 as function 
of $\eta_{\rm s}$.
In figure 2 the functional forms of $u_{z,(n,s)}^{(0)}$ of $n=1$, 2, and 3 are shown for 
some values of $\eta_{\rm s}$.
It is noted that in the limit of $\eta_{\rm s}=\infty$, $K_{n,{\rm s}}$ and $u_{z,(n,s)}^{(0)}$ tend, 
respectively, to
\begin{equation}
      K_{n,{\rm s}}=n-1, \quad \rho_0 u_{z,(n,s)}^{(0)}\propto {\rm exp}(-\eta^2/2){\cal H}_{n-1}(\eta),
\label{}
\end{equation}
which are the same as the eigen-value and eigen-function of non-terminated isothermal disks 
(Okazaki et al. 1987), where ${\cal H}_{n-1}(\eta)$ is the Hermite polynomial of the order of $n-1$ 
with argument $\eta$.

\begin{figure}
\begin{center}
    \FigureFile(80mm,80mm){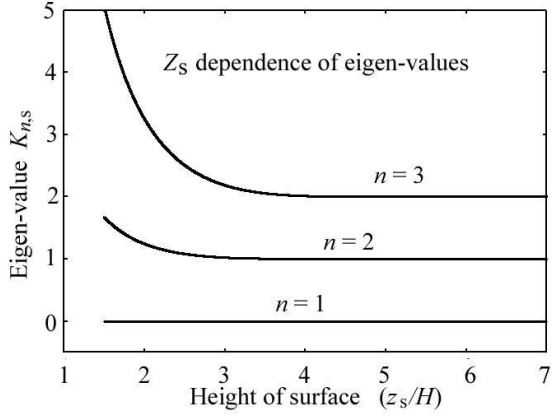}
\end{center}
\caption{
Eigen-value, $K_{n,{\rm s}}$, of purely vertical oscillations as functions of cutting  
height, $\eta_{\rm s}(\equiv z_{\rm s}/H)$.
Three modes of $n=1$, $n=2$, and $n=3$ are considered.
In the limit of $\eta_{\rm s}=\infty$, $K_{n,{\rm s}}$ tends to that of the non-cutted isothermal
disks, which is $n-1$, i.e., $K_{n,{\rm s}}$ tends to $K_{n,{\rm s}}=n-1$.
It is noted that when $n=1$, $K_{n,{\rm s}}=0$, free from s. 
}
\end{figure}
\begin{figure}
\begin{center}
    \FigureFile(80mm,80mm){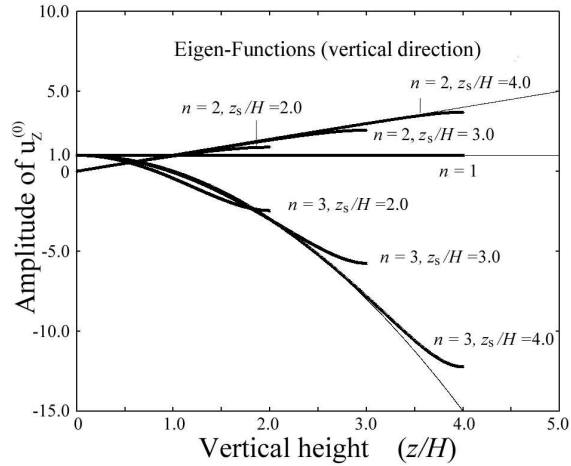}
\end{center}
\caption{
Eigen-functions, $u_{z,(n,{\rm s})}^{(0)}$, of purely vertical oscillations in disks of finite thickness.
Eigen-functions of three modes ($n=1$, $n=2$, and $n=3$) are shown for three disks
of $\eta_{\rm s}=$2.0, 3.0, and 4.0.
The amplitude of oscillations are taken arbitrary.
It is noted that in the limit of non-terminated disks ($\eta_{\rm s}=\infty$), the eigen-function of
$n=2$ tends to  $u_{z,(n,{\rm s})}^{(0)}(\eta)=\eta$, and that of $n=3$ to $u_{z,(n,{\rm s})}^{(0)}
(\eta)=1-\eta^2$.
In the case of $n=1$, $u_{z,(n,{\rm s})}^{(0)}$ is unity, independent of $\eta_{\rm s}$.
Eigen-fuctions in the non-terminated isothermal disks are shown by thin curves. 
} 
\end{figure}

\subsection{Orthogonality of Eigen-functions}

The orthogonality of eigen-functions, $u_{z,(n,s)}^{(0)}$, of different $n$'s is noted here, 
since we need it to derive wave equation in the radial direction, as is shown in the subsequent section.
Now, $u_{z,(n,s)}^{(0)}(r,\eta)$ is separated into $r$- and $\eta$-dependent terms as
$u_{z,(n,s)}^{(0)}(r,\eta)=f(r)g_{n,s}^{(0)}(\eta)$.
At the present stage, the functional form of $f(r)$ is arbitrary, which is determined later.
The eigen-function $g_{n,s}^{(0)}$ then satisfies the following wave equation:
\begin{equation}
     \frac{\partial}{\partial \eta}\biggr[{\rm exp}(-\eta^2/2)\frac{\partial
        g_{n,s}^{(0)}}{\partial \eta}\biggr]
        +{\rm exp}(-\eta^2/2)K_{n,s}g_{n,s}^{(0)}=0
\label{}
\end{equation}
[see equations (\ref{zeroth}) and (\ref{eigenvalue})].
Hence, if the above equation is multiplied by $g_{n',s}^{(0)}$ and integrated from $-\eta_s$ to $\eta_s$,
we have, after integrating by part with the help of boundary condition (\ref{boundary3}), 
\begin{equation}
  (K_{n,s}-K_{n',s})\int_{-\eta_s}^{\eta_s}{\rm exp}(-\eta^2/2)
        g_{n,s}^{(0)}g_{n',s}^{(0)}d\eta=0.
\label{orthogonality0}
\end{equation}
This equation gives the orthogonality relation:
\begin{equation}
        \langle g_{n,s}^{(0)}(\eta),\ g_{n',s}^{(0)}(\eta)\rangle = 0, \quad {\rm for} \quad n'\not= n.
\label{orthogonality}
\end{equation}
Here and hereafter, $\langle a,\ b\rangle$ means the integration of the product of
$a(\eta)$ and $b(\eta)$ over $\eta=-\eta_{\rm s}$ to $\eta=\eta_{\rm s}$ with weighting function 
${\rm exp}(-\eta^2/2)$.

\section{Wave Equation in Radial Direction}

Our next object is to proceed to the next order of approximations of equation (\ref{wave}), 
where $u_r$ is taken into account.
Then, $u_z$ is no longer separated into a product of $\eta$- and $r$-dependent terms.
That is, we expand $u_{z,({\rm n,s})}$ as
\begin{eqnarray}
     u_z(r,\eta)&&=u_{z,(n,s)}^{(0)} + u_{z,(n,s)}^{(1)}+...   \nonumber \\
                &&\equiv f(r)[g_{n,s}^{(0)}(\eta)+g_{n,s}^{(1)}(r,\eta)+....].
\label{expansion}
\end{eqnarray}
In the lowest order of approximations, $u_z(r,\eta)$ is separated as $f(r)g_{n,s}^{(0)}(\eta)$,
but in the next order quantities, $u_z(r,\eta)$ is a product of $f(r)$ and a weakly
$r$-dependent term $g_{n,s}^{(1)}(r,\eta)$.
Then, from equation (\ref{wave}), as equation describing $fg_{n,s}^{(1)}$,
we have 
\begin{equation}
      f(r)\biggr(\frac{\partial^2}{\partial \eta^2}
                 -\eta\frac{\partial}{\partial \eta}+K_{n,s}\biggr)g_{n,s}^{(1)}(r,\eta)
            =-\epsilon(r)f(r)g_{n,s}^{(0)}(\eta)
             -H\biggr[\frac{\partial}{\partial \eta}-\frac{c_{\rm A}^2/2}{c_{\rm s}^2+c_{\rm A}^2}\eta\biggr]
            \frac{\partial u_r^{(0)}}{\partial r},
\label{perturbation}            
\end{equation}
where the superscript (0) has been attached to $u_r^{(0)}$ in order to emphasize that $u_r$ in its lowest 
order quantity is enough in this equation.

The next subject is to express $u_r^{(0)}$ in terms of $u_z^{(0)}$ in order to solve equation 
(\ref{perturbation}).
By the same procedures as in paper I, we have (see paper I), after some manipulations, 
\begin{equation}
     u_r^{(0)}={\cal L}_{\rm s}(f)\biggr(\frac{d}{d \eta}-\eta\biggr)g_{n,s}^{(0)}(\eta)
               +{\cal L}_{\rm A}(f)\biggr(\frac{d}{d\eta}-\frac{1}{2}\eta\biggr)g_{n,s}^{(0)}(\eta),
\label{expression-ur}
\end{equation}
where ${\cal L}_{\rm s}$ and ${\cal L}_{\rm A}$ are operators defined by 
\begin{equation}
     {\cal L}_{\rm s}(f)=\frac{c_{\rm s}^2/H}{-(\omega-m\Omega)^2+\kappa^2}
           \biggr[\frac{d}{d r}
           -\frac{d{\rm ln}(\omega-m\Omega)}{d r}\biggr]f(r)
\label{Ls}
\end{equation}
and
\begin{equation}
     {\cal L}_{\rm A}(f)=\frac{c_{\rm A}^2/H}{-(\omega-m\Omega)^2+\kappa^2}
           \biggr[\frac{d}{d r}
           -\frac{d{\rm ln}(\omega-m\Omega)}{d r}\biggr]f(r).
\label{LA}
\end{equation}

Equation (\ref{perturbation}) is an inhomogeneous differential equation of $g_{n,s}^{(1)}(r,\eta)$
with respect to $\eta$.
Hence, the solvability condition of inhomogeneous differential equation (\ref{perturbation}) 
with respect to $\eta$ 
is that when the right-hand side of equation (\ref{perturbation}) is expanded by the orthogonal set of
functions, $g_{n',s}^{(0)}(\eta)$, it does not contain any term proportional to $g_{n,s}^{(0)}(\eta)$.
That is,  the condition is 
\begin{equation}
     \epsilon f\langle g_{n,s}^{(0)},g_{n,s}^{(0)}\rangle
     + AH\frac{d}{dr}{\cal L}_{\rm s}(f)
     + BH\frac{d}{dr}{\cal L}_{\rm A}(f)=0,
\label{wave-equation0}
\end{equation}
where
\begin{equation}
  A=\biggr\langle g_{n,s}^{(0)}(\eta),
         \biggr(\frac{d}{d \eta}
         -\frac{1}{2}\frac{c_{\rm A}^2}{c_{\rm s}^2+c_{\rm A}^2}\eta\biggr)
         \biggr(\frac{d}{d\eta}-\eta\biggr) g_{n,s}^{(0)}(\eta)\biggr\rangle 
\label{AA}
\end{equation}
and
\begin{equation}
   B=\biggr\langle g_{n,s}^{(0)}(\eta),\ 
         \biggr(\frac{d}{d \eta}
         -\frac{1}{2}\frac{c_{\rm A}^2}{c_{\rm s}^2+c_{\rm A}^2}\eta\biggr)
         \biggr(\frac{d}{d\eta}-\frac{1}{2}\eta\biggr)g_{n,s}^{(0)}(\eta)\biggr\rangle.
\label{B}
\end{equation}
This solvability condition (\ref{wave-equation0}) is an ordinary differential equation of $f(r)$ 
with respect to $r$, 
since $\eta$ disappears here by integrations over $\eta$.

If expressions for ${\cal L}_{\rm s}(f)$ and ${\cal L}_{\rm A}(f)$ given by
equations (\ref{Ls}) and (\ref{LA}) are substituted into equation (\ref{wave-equation0}) 
under approximations of
neglecting weak radial dependences of $H$, $c_{\rm s}$, and $c_{\rm A}$, we see that equation 
(\ref{wave-equation0}) is a wave equation describing behavoir of $f(r)$ in the radial direction
and is expressed as
\begin{equation}
    -\frac{Ac_{\rm s}^2+Bc_{\rm A}^2}{\langle g_{n,s}^{(0)}, g_{n,s}^{(0)}\rangle}
   \frac{d}{dr}\biggr[\frac{\omega-m\Omega}{(\omega-m\Omega)^2-\kappa^2}\frac{d}{dr}
      \biggr(\frac{f}{\omega-m\Omega}\biggr)\biggr]
          +\epsilon f=0.
\label{radial-eq1}
\end{equation}
To obtain detailed expressions for $A$ and $B$ from equations (\ref{AA}) and (\ref{B}), 
we need some manipulations, which are given in appendix.
The results show that
\begin{equation}
   A= \biggr[-(K_{n,s}+1)+\frac{1}{4}\frac{c_{\rm A}^2}{c_{\rm s}^2+c_{\rm A}^2}\biggr]I_{n,s,0}
      +\frac{1}{4}\frac{c_{\rm A}^2}{c_{\rm s}^2+c_{\rm A}^2}I_{n,s,2}
      -\frac{1}{2}\frac{c_{\rm A}^2}{c_{\rm s}^2+c_{\rm A}^2}S
\label{equationA}
\end{equation}
and
\begin{equation}
       B=\biggr[-(K_{n,s}+\frac{3}{4})+\frac{1}{4}\frac{c_{\rm A}^2}{c_{\rm s}^2+c_{\rm A}^2}\biggr]I_{n,s,0}
      +\frac{1}{4}I_{n,s,2}
      +\frac{1}{2}\frac{c_{\rm s}^2}{c_{\rm s}^2+c_{\rm A}^2}S
\label{equationB}
\end{equation}
where
\begin{equation}
     I_{n,s,0}= \langle g_{n,s}^{(0)}(\eta),\ g_{n,s}^{(0)}(\eta)\rangle \quad {\rm and}\quad
     I_{n,s,2}= \langle \eta g_{n,s}^{(0)}(\eta), \ \eta g_{n,s}^{(0)}(\eta)\rangle,
\label{Ins0-Ins2}
\end{equation}
and $S$ is a surface value defined by
\begin{equation}
       S=\eta_{\rm s}{\rm exp}(-\eta_s^2/2)[g_{n,s}^{(0)}(\eta_s)]^2.
\label{surface}
\end{equation}
Then, $Ac_{\rm s}^2+Bc_{\rm A}^2$ becomes
\begin{equation}
   Ac_{\rm s}^2+Bc_{\rm A}^2=-\biggr[(c_{\rm s}^2+c_{\rm A}^2)K_{n,s}
        +\biggr(c_{\rm s}^2+\frac{1}{2}c_{\rm A}^2\biggr)\biggr]I_{n,s,0}
        +\frac{1}{2}c_{\rm A}^2\frac{c_{\rm s}^2+c_{\rm A}^2/2}{c_{\rm s}^2+c_{\rm A}^2}I_{n,s,2}.
\end{equation}   

By using expressions for $Ac_{\rm s}^2+Bc_{\rm A}^2$ given above, 
we can finally write down the wave equation (\ref{radial-eq1}) in the form
\begin{equation}
       A_{n,s}\biggr(c_{\rm s}^2+\frac{1}{2}c_{\rm A}^2\biggr)
   \frac{d}{dr}\biggr[\frac{\omega-m\Omega}{(\omega-m\Omega)^2-\kappa^2}\frac{d}{dr}
      \biggr(\frac{f}{\omega-m\Omega}\biggr)\biggr]
          +\epsilon f=0,    
\label{wave-eq-radial}
\end{equation}
where $A_{n,{\rm s}}$ is defined by
\begin{equation}
    A_{n,{\rm s}}=\frac{c_{\rm s}^2+c_{\rm A}^2}{c_{\rm s}^2+c_{\rm A}^2/2}K_{n,s}
     +1-\frac{c_{\rm A}^2/2}{c_{\rm s}^2+c_{\rm A}^2}\frac{I_{n,s,2}}{I_{n,s,0}},
\label{A}
\end{equation}
where the ratio $I_{n,s,2}/I_{n,s,0}$ can be obtained by using vertical eigenfunctions
shown in figure 2.
It is noted that in the limit of $\eta_s=\infty$, $I_{n,s,2}/I_{n,s,0}$ goes to $2n-1$.
Then, $A_{n,s}$ given by equation (\ref{A}) tends to $nA$ in paper I, and equation (\ref{radial-eq1}) 
becomes equal to equation (36) in paper I.

In paper I we have introduced a new unknown function ${\tilde f}$ defined by
${\tilde f}=f/(\omega-m\Omega)$.
By using this function, we can reduce equation (\ref{radial-eq1}) to
\begin{equation}
     \frac{1}{\omega-m\Omega}\frac{d}{dr}\biggr[\frac{\omega-m\Omega}
        {(\omega-m\Omega)^2-\kappa^2}\frac{d {\tilde f}}{dr}\biggr]   
        +\frac{\epsilon}{A_{n,{\rm s}}\Omega_\bot^2H^2}{\tilde f}=0.
\label{radial-eq2}
\end{equation}

\subsection{Radial Eigenvalue Problems}

Next, we solve equation (\ref{radial-eq2}) as an eigen-value problem to study where the oscillations
are trapped and how much the eigen-frequency of the trapped oscillations are.
We introduce now a new independent variable $\tau(r)$ defined by
\begin{equation}
     \tau(r)=\int_{r_{\rm i}}^r\frac{{\tilde \omega}^2(r')-\kappa^2(r')}{-{\tilde \omega}(r')}dr', 
           \quad \tau_{\rm c}\equiv\tau(r_{\rm c}),
\label{3.19}
\end{equation}
where ${\tilde \omega}$ is defined by ${\tilde \omega}=(\omega-m\Omega)$,
and $r_{\rm i}$ is the inner edge of disks where a boundary condition is imposed.
Then, equation (\ref{radial-eq2}) is written in the form:
\begin{equation}
      \frac{d^2{\tilde f}}{d\tau^2} +Q {\tilde f}=0,
\label{3.20}
\end{equation}
where
\begin{equation}
      Q(\tau)=\frac{{\tilde\omega}^2}{{\tilde\omega}^2-\kappa^2}
          \frac{\epsilon}{A_{n,{\rm s}}\Omega_\bot^2H^2}.
\label{3.22}
\end{equation}
Equations (\ref{3.20}) and (\ref{3.22}) show that the propagation region of oscillations is the region 
where $Q>0$, which is the region of $\epsilon >0$.

We solved equation (\ref{3.20}) by a standard WKB method with relevant boundary conditions
(for details, see Silbergleit et al. 2001 and Ortega-Rodrigues et al. 2008).
The WKB approximation shows that the solution of equation (\ref{3.20}) can be represented as
\begin{equation}
    {\tilde f}\propto Q^{-1/4}(\tau){\rm cos}\ [\Phi(\tau)-\Phi_{\rm c}]
\label{3.21}
\end{equation}
in the whole capture region $0<\tau < \tau_{\rm c}$, except small vicinities of its boundaries
of $\tau=0$ (i.e., $r=r_{\rm i}$) and $\tau=\tau_{\rm c}$ (i.e., $r_{\rm c}$).
Here, $\Phi(\tau)$ is defined by
\begin{equation}
     \Phi(\tau)=\int_0^\tau Q^{1/2}(\tau')d\tau'=\int_{r_{\rm i}}^rQ^{1/2}(r')
          \frac{{\tilde \omega}^2(r')-\kappa^2(r')}{-{\tilde\omega}(r')}dr',
\label{3.23}
\end{equation}
and $\Phi_{\rm c}$ is a constant to be determined by boundary conditions.
To determine the outer boundary condition, we take into account the fact that 
the capture radius, $r_{\rm c}$, 
is a turning point of equation (\ref{3.20}) since the sign of $\epsilon$ changes there.
The inner boundary condition we adopted is 
${\tilde f}=0$ at $r=r_{\rm i}$.
As the inner boundary radius we take the marginary stable radius.
Then, WKB analyses show that the trapping condition is\footnote{see Silbergleit et al. (2001)
or paper I for the case where boundary condition $d{\tilde f}/dr=0$ is adopted at $r=r_{\rm i}$.}
\begin{equation}
    \int_0^{\tau_{\rm c}}Q^{1/2}d\tau=\pi(n_r+3/4), 
\label{3.24}    
\end{equation}
where $n_r(=0,1,2,...)$ is zero or a positive integer specifying the node number of 
${\tilde f}$ in the radial direction.
The constant $\Phi_{\rm c}$ is determined as
\begin{equation}
    \Phi_{\rm c}=\pi/2.
\end{equation}

For a given set of parameters, including spin parameter $a_*$ 
and mass of neutron stars, $M$, any solution of equation (\ref{3.24}) specifies $r_{\rm c}$,
which gives $\omega$ of the trapped oscillation through equation (\ref{eigenvalue}).
In other words, $\omega$ and $r_{\rm c}$ are related by equation (\ref{eigenvalue}), i.e., 
$\omega=\omega(r_{\rm c})$ or $r_{\rm c}=r_{\rm c}(\omega)$, and
the trapping condition determines $r_{\rm c}$ or $\omega$ (see also the next section and
figure 3).

\section{Numerical Results}

We consider the same disks as in paper I, except that the present disks are terminated 
at a certain height, $z_{\rm s}(\equiv\eta_{\rm s}H)$.
The height is taken as a parameter, independent of $r$.
The temperature distribution adopted in the radial direction is that of the standard disks where
the gas pressure dominates over the radiation pressure and the opacity mainly comes from
the free-free processes, i.e., the acoustic speed on the equator, $c_{\rm s0}$, is taken as 
(e.g., Kato et al. 2008)
\begin{equation}
     c_{{\rm s}0}^2=1.83\times 10^{16}\biggr(\alpha \frac{M}{M_\odot}\biggr)^{-1/5}
         \biggr(\frac{\dot M}{{\dot M}_{\rm crit}}\biggr)^{2/5}\biggr(\frac{r}{r_{\rm g}}\biggr)^{-9/10}
         \ {\rm cm}^2\ {\rm s}^{-2},
\label{4.1}
\end{equation}
where $\alpha$ is the conventional viscosity parameter, $r_{\rm g}$ is the Schwarzschild radius defined by 
$r_{\rm g}=2GM/c^2$, 
and ${\dot M}_{\rm crit}$ is the critical mass-flow
rate defined by the Eddington luminosity.
Throughout this paper, we fix $\alpha=0.1$, ${\dot M}/{\dot M}_{\rm crit}=0.3$, and $M/M_\odot=2.0$,
as in paper I, since our purpose here is to examine how disk termination at a certain height affects
on frequencies of trapped oscillations.

\begin{figure}
\begin{center}
    \FigureFile(80mm,80mm){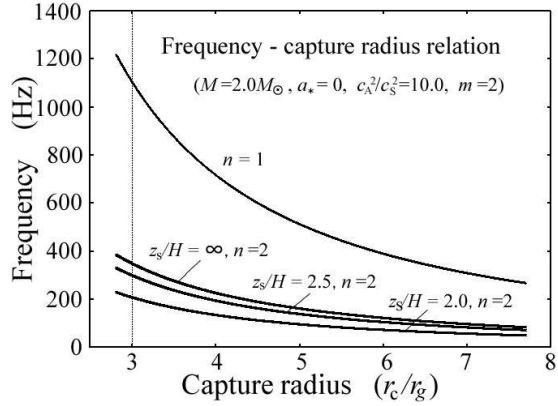}
\end{center}
\caption{
Frequency $\omega$ -- capture radius $r_{\rm c}$ relation for $n=1$ and $n=2$ oscillations
in the case of $a_*=0$.
Parameter values adopted are $M=2.0M_\odot$ and $c_{\rm A}^2/c_{\rm s}^2=10.0$.
In the case of $n=1$ the frequency -- capture radius relation is free from $\eta_{\rm s}(\equiv z_{\rm s}/H)$.
In the case of $n=2$, however, the relation depends on $\eta_{\rm s}$.
As the normalized disk thickness decreases the curve moves downward on the frequency -- radius diagram.
} 
\end{figure}

Before examinig parameter dependences of frequency, $\omega$, of trapped oscillations,
it is helpful to study how $\omega$ is related to the capture radius $r_{\rm c}$.
The $\omega$ -- $r_{\rm c}$ relation is specified by $\epsilon=0$ and given by
equation (\ref{eigenvalue}), i.e., 
\begin{equation}
  \omega=\biggr[m\Omega-\biggr(\frac{c_{\rm s}^2+c_{\rm A}^2}{c_{\rm s}^2+c_{\rm A}^2/2}
         K_{n,s}+1\biggr)\Omega_\bot\biggr]_{\rm c}.
\label{acoustic}
\end{equation}
The $\omega$ -- $r_{\rm c}$ relation depends on $m$, $c_{\rm A}^2/c_{\rm s}^2$ and $\eta_{\rm s}$ 
in addition to the node number $n$ in the vertical direction.
As mentioned before, we consider only the case of $m=2$ throughout this paper.
Figure 3 shows relation (\ref{acoustic}) to oscillation modes of $n=1$ and $n=2$, 
in cases of disks with $\eta_{\rm s}=2.5$ and 2.0. 
The strength of magnetic fields is taken as $c_{\rm A}^2/c_{\rm s}^2=10.0$.
For comparison, the case of $\eta_{\rm s}=\infty$ is also shown.
It is noted that in the case of $n=1$, the $\omega$ -- $r_{\rm c}$ relation is free from 
$\eta_{\rm s}$.
This comes from the fact that $K_{n,s}=0$ when $n=1$.
In the case of $n\not= 1$, however, the $\omega$ -- $r_{\rm c}$ relation is affected 
by $\eta_{\rm s}$ and moves downward on the frequency-radius diagram as the height of termination decreases.
The downward shifts of the $\omega$ -- $r_{\rm c}$ curve on the frequency -- radius diagram
means that the frequency of trapped oscillations decreases and the trapped region extends in the
radial direction, although detailed values of the frequencies are obtained after
the wave equation (\ref{radial-eq2}) is solved.  

\begin{figure}
\begin{center}
    \FigureFile(80mm,80mm){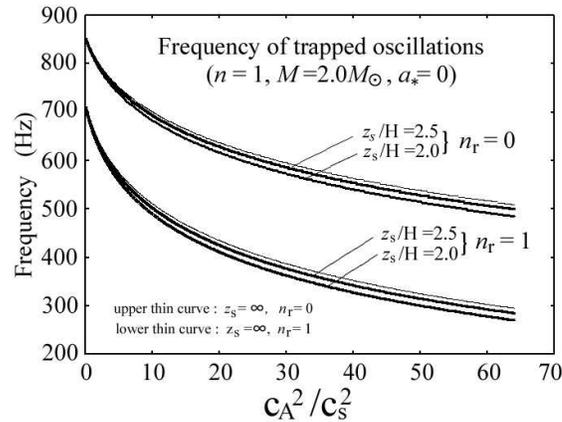}
\end{center}
\caption{
Frequency-$c_{\rm A}^2/c_{\rm s}^2$ relation of trapped oscillations of $n=1$.
Two cases of $n_r=0$ and $n_r=1$ are shown for two disks ($\eta_{\rm s}=2.0$ and 2.5).
For comparison, the cases of disks with $\eta_{\rm s}=\infty$ are shown by thin curves. 
It is noted that the relation depends little on height of termination, compared with the cases of $n=2$
shown in figure 5.
} 
\end{figure}
\begin{figure}
\begin{center}
    \FigureFile(80mm,80mm){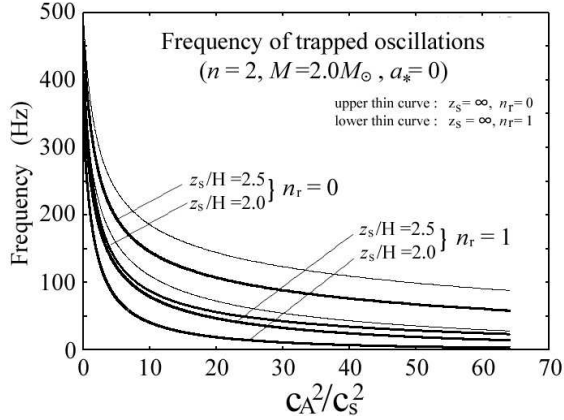}
\end{center}
\caption{
Frequency-$c_{\rm A}^2/c_{\rm s}^2$ relation for trapped oscillations of $n=2$.
Two cases of $n_r=0$ and $n_r=1$ are shown for two disks ($\eta_{\rm s}=2.0$ and
$\eta_{\rm s}=2.5$).
For comparison, the cases of $\eta_{\rm s}=\infty$ are shown by thin curves.
It is noted that the frequency of trapped oscillations decreases as $\eta_{\rm s}(\equiv z_{\rm s}/H)$ 
decreases. 
} 
\end{figure}

The frequencies of trapped oscillations obtained by solving equation (\ref{radial-eq2}) by
the WKB method of the last section are shown in figures 4 -- 6.
In figure 4, the frequencies of trapped $n=1$ oscillations with $n_{\rm r}=0$ or $n_{\rm r}=1$
are shown as functions of $c_{\rm A}^2/c_{\rm s}^2$ for three cases of $\eta_{\rm s}=\infty$,
2.5, and 2.0.
Figure 5 is the same as figure 4, except that the $n=2$ oscillations are considered.
Comparison of figures 4 and 5 shows that in the case of oscillations of $n=2$,
frequency of trapped oscillations are much affected by termination of disks,
i.e., their frequencies decrease with decrease of disk thickness (decrease of $\eta_{\rm s}$).
Figure 6 shows effects 
of spin of the central source on frequency of trapped oscillations
in the case of $\eta_{\rm s}= 2.0$ for oscillations of
$n=1$ and 2 with $n_r=0$ and 1.
The value of $c_{\rm A}^2/c_{\rm s}^2$ adopted is 10.0.
As in the case of non-terminated disks, the spin acts so as to increase the
frequencies of trapped oscillations.

\begin{figure}
\begin{center}
    \FigureFile(80mm,80mm){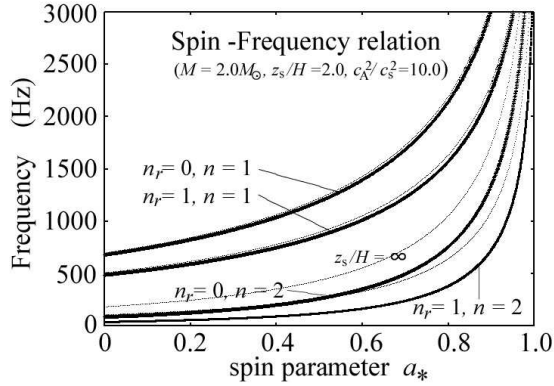}
\end{center}
\caption{
Frequency-spin relation for oscillations of $n=1$ and 2 with two different $n_{\rm r}$.
The disk thickness is taken as $\eta_{\rm s}=2.0$, and $c_{\rm A}^2/c_{\rm s}^2$ 
adopted is 10.0.
The thin four curves show the cases of $\eta_{\rm s}=\infty$.
Their upper two curves are for $n_{\rm r}=0$ and $n_{\rm r}=1$ both with $n=1$, 
which are almost overlapped with those of $\eta_{\rm s}=2.0$.
The lower two curves among four thin ones are for $n_{\rm r}=0$ ($n=2$) and $n_{\rm r}=1$ ($n=2$) from the upper.
} 
\end{figure}

Next, parameter dependences of the radial width, $r_{\rm c}-r_{\rm i}$, of the trapped region 
(i.e., capture zone) are examined for some modes of oscillations.
Figure 7 shows the $c_{\rm A}^2/c_{\rm s}^2$-dependence of the width for four oscillation modes
($n_r=0$ and $n_r=1$ with $n=1$, and $n_r=0$ and $n_r=1$ with $n=2$)
in the case where $\eta_{\rm s}=2.0$ and $a_*=0$.
For comparison, the cases of $\eta_{\rm s}=\infty$ and $a_*=0$ are shown by thin curves for the
above four oscillation modes.
In general, the width of trapped region increases with decrease of disk thickness.
This trend is prominent in oscillations of $n=2$, although it is not so in the oscillations of $n=1$.
In the case of highly rotating sources, however, the trapped region decreases with increase of $a_*$ 
as shown in figure 8, where $c_{\rm A}^2/c_{\rm s}^2$ is again fixed to $c_{\rm A}^2/c_{\rm s}^2=10.0$.

\begin{figure}
\begin{center}
    \FigureFile(80mm,80mm){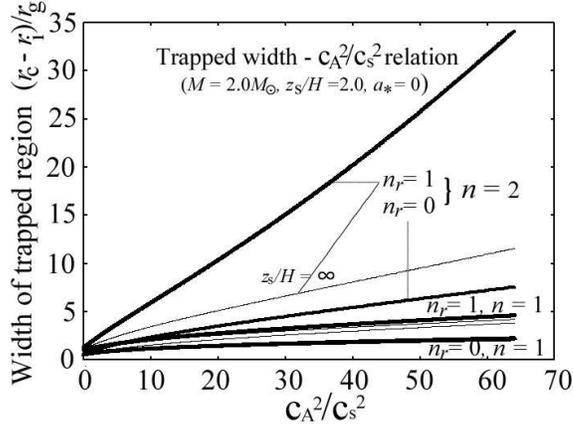}
\end{center}
\caption{
Width of trapped region, $r_{\rm c}-r_{\rm i}$, as functions of $c_{\rm A}^2/c_{\rm s}^2$ for
four oscillation modes of $n_r=0$ and 1 both with $n=1$, and $n_r=0$ and 1 both with $n=2$
in disks with $\eta_{\rm s}=2.0$.
Other parameters adopted are $M=2.0M_\odot$ and $a_*=0$.
For comparison, the corresponding cases of $\eta_{\rm s}=\infty$ are shown by thin curves.
The curves of $\eta_{\rm s}=2.0$ and $\eta_{\rm s}=\infty$ cannot be distinguished in the figure
when $n=1$ and $n_r=0$.
The curves of $\eta_{\rm s}=\infty$ in the cases of $n_{\rm r}=0$ (with $n=2$) and of $n_{\rm r}=1$
(with $n=1$) are just below the curve of $\eta_{\rm s}=2.0$ with $n_{\rm r}=1$ and $n=1$.
It should be noted that termination of disks in the vertical direction makes the width of trapped
region wider especially in cases of oscillations of $n=2$.   
} 
\end{figure}
\begin{figure}
\begin{center}
    \FigureFile(80mm,80mm){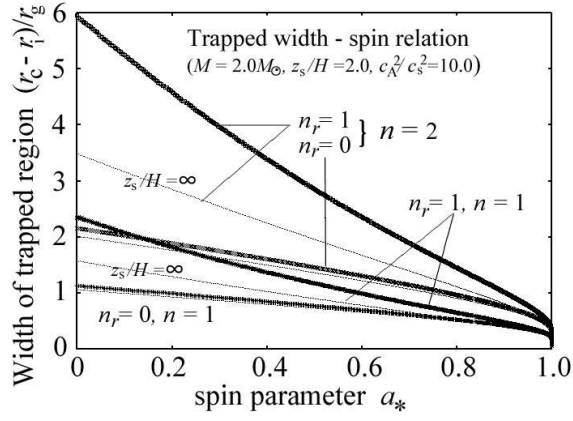}
\end{center}
\caption{
Width of trapped region as functions of $a_*$ for four oscillation modes of
$n_r=0$ and 1 both with $n=1$, and of $n_r=0$ and 1 both with $n=2$ in disks with $\eta_{\rm s}=2.0$.
Other parameters adopted are $M=2.0M_\odot$ and $c_{\rm A}^2/c_{\rm s}^2=10.0$.
For comparison, the corresponding curves in the case of $\eta_{\rm s}=\infty$ are shown by
thin curves.
The curves of $\eta_{\rm s}=\infty$ and $\eta_{\rm s}=2.5$ cannot be distinguished in the figure
when $n_{\rm r}=0$ with $n=1$.

} 
\end{figure}

Finally, validity of the perturbation method adopted in this paper is checked.
We have separated the $z$- and $r$- dependences of eigen-functions by assuming that 
the dimensionless quantity $\epsilon(r)$ introduced by equation (\ref{eigenvalue0}) is a small positive quantity 
in the radial propagation region of oscillations, i.e., $0<\epsilon(r)< 1$ in $r_{\rm i}<r<r_{\rm c}$.
This assumption can be checked after the radial eigen-value problem has been solved.
The results show that the approximation is allowed in the first order of approximation.
As an example, the radial distribution of $\epsilon(r)$ obtained in some cases of
$\eta_{\rm s}=2.0$ and 2.5 are shown in figure 9.

\begin{figure}
\begin{center}
    \FigureFile(80mm,80mm){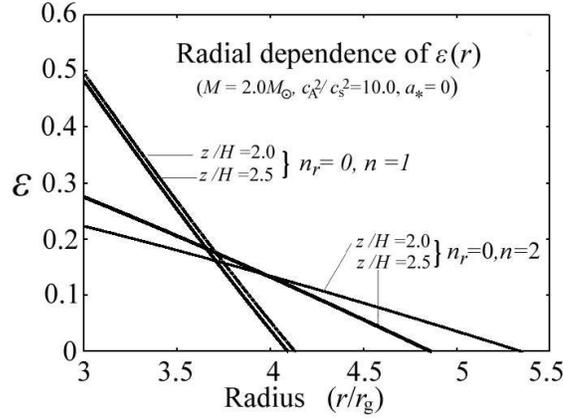}
\end{center}
\caption{
Radial dependence of $\epsilon(r)$.
The dependence is shown for two modes of oscillations of $n=1$ and $n=2$ in cases
where the disks are terminated at $z/H=2.0$ and $z/H=2.5$.
The value of $\epsilon$ vanishes at the capture radius $r_{\rm c}$.
The region of $r< r_{\rm c}$ is the propagation region of oscillations.
The requirement of $0<\epsilon<1$ in the propagation region is satisfied. 
} 
\end{figure}

\section{Discussion}

In this paper we have examined trapping of two-armed $(m=2)$ vertical p-mode 
oscillations in isothermal disks which are terminated at a certain height and subject to toroidal magnetic
fields.
The two-armed vertical p-mode oscillations are of interest since they
are trapped in the innermost region of disks with moderate frequencies because of the very
fact of $m=2$ (Kato 2010).
For the vertical p-mode oscillations to be a good candidate of observed QPOs, their
frequencies should not be robust but vary with time with a change of disk structure.
Furthermore, the time variations of their different modes must correlate in such a way as
they can describe the correlated variations of observed QPOs.

In the present model of QPOs, the major causes of time variation of frequencies of trapped oscillations
are time changes of
i) toroidal magnetic fields, ii) disk temperature, and iii) vertical thickness of disks.
Time change of toroidal magnetic fields is generally expected, since magnetic fields are 
wound by differential rotation and loosened by magnetic reconnection.
Time change of disk temperature will occur if, for example, mass accretion rate varies.
Time change of vertical thickness of cool disks will also occur in real disks, since
cool disks with strong toroidal magnetic fields and corona are expected 
as disks which bridge between ADAFs and optically thick disks 
(Machida et al. 2004, Oda et al. 2007, 2009, 2010).
Evaporation of disk gas to hot corona will also depend on stages of disk activity.

The results of this paper show that a change of vertical disk thickness on frequencies of trapped
oscillations is little in the case of $n=1$ oscillations.
This is related to the fact that the eigen-value of purely vertical p-mode oscillations, $K_{n,{\rm s}}$, 
is independent of $\eta_{\rm s}$, and is zero when $n=1$, i.e., $K_{1,{\rm s}}=0$.
In the case of oscillatios of $n=2$, however, vertical termination of disk thickness has non-negligible 
effects on their frequencies.
That is, as the height decreases, the frequencies of trapped oscillations of $n=2$ decrease 
and the radial extend of trapping region becomes wide.
The trapped region is, however, not wide compared with that of c-mode oscillations 
(Silbergleit et al 2001) in the case of small $a_*$.
Different from c-mode oscillations which are incompressible motions in the limit of $a_*=0$, 
the two-armed vertical p-mode oscillations are the fast mode of the three MHD modes, 
and thus strong density and temperature variations are associated with the oscillations.
Hence, in the case of disks which are vertically terminated by hot coronae, hard X-ray time variations by 
inverse Compton processes will be expected as the results of geometrically thin disk oscillations.

In the case of $n=2$ oscillations, the trapped region is wide compared with the case of $n=1$ oscillations,
as mentioned above.
Thus, our basic approximation introduced in this paper that the radial variations of 
$c_{\rm A}^2/c_{\rm s}^2$ and $\eta_{\rm s}$ are negligible should be improved in more realistic
studies.
Furthermore, more careful considerations on inner boundary condition will be 
necessary to do more quantitative studies.
Concerning the radius where an inner boundary condition is imposed, however, the ISCO will be relevant  
as the first approximation, unless strong poloidal magnetic fields anchored to the central sources 
are present. 
 
One of basic assumptions involved in our analyses is that the disks are vertically isothermal
and $c_{\rm A}^2$ is also constant in the vertical direction.
We think that the assumption of vertical isothermality will be better than that of 
polytropic disks where temperature decreases from equator toward disk surface,\footnote{
Trapping of two-armed p-mode oscillations in polytropic disks has been examined (Kato 2010).
The results show that an increase of polytropic index decreases the frequency of the
trapped oscillations.
}
especially in disks surrounded by hot corona.
Examinations how our present results are modified if $c_{\rm A}^2$ is not constant in the 
vertical direction are of importance to evaluate whether the present disk oscillation model is relevant to 
describe the observed QPOs.
We suppose that the frequency of trapped oscillations decreases in the case where $c_{\rm A}^2$ increases
toward surface, compared with the case where $c_{\rm A}^2$ is constant in the vertical direction
(other parameters are retained unchanged).
This supposition is based on the following considerations.
At the capture radius $r_{\rm c}$, the trapped oscillations are roughly vertical.
That is, at the capture radius the frequency of the trapped oscillations in the corotating frame, 
$\omega-m\Omega$, is the frequency of the fast mode of the three MHD waves propagating in the vertical direction
(at the present stage, the capture radius is still unknown).
This means that when $c_{\rm A}^2$ increases with $z$, $(\omega-m\Omega_{\rm c})^2$ is larger than 
that in the case of $c_{\rm A}^2=$ const.
Then, since we are considering here the oscillations which are inside of the inner Lindblad resonance (i.e., 
$r<r_{\rm c}<r_{\rm IL}$), the curve describing the $\omega$ -- $r_{\rm c}$ relation runs below the 
corresponding curve of $c_{\rm A}^2=$ const.
on the frequency-radius diagram (see figure 3).
Next, we consider trapping of the oscillations in the radial direction by using the WKB method.
An increase of vertically averaged $c_{\rm A}^2$ may lead to decrease of $\omega$ of the trapped oscillations
[see $c_{\rm A}^2$ dependence of $Q$ given by equation (\ref{3.22})], 
even if the $\omega$ -- $r_{\rm c}$ relation would be unchanged.
Together with this and the downward shift of the $\omega$ -- $r_{\rm c}$ relation mentioned above,
an increase of $c_{\rm A}^2$ towards the surface will decrease 
$\omega$ of the trapped oscillations.
This is a supposition based on rough considerations, and more careful examinations are necessary to do
a quantitative estimate.
It is noted, however, that we suppose that the curve of the frequency correlation between $n_{\rm r}=0$ 
and $n_{\rm r}=1$ oscillations (both with $n=1$) does not differ much from that in the case of $c_{\rm A}^2$ being constant
in the vertical direction, 
since frequencies of both oscillations shift in the same direction on the
frequency -- frequency diagram (see studies concerning effects of change of $\eta_{\rm s}$ 
on the correlation curve, which are given in the subsequent paper).

In previous studies on two-armed vertical p-mode oscillations in infinitely extended isothermal disks
($\eta_{\rm s}=\infty$),
we showed that the observed frequency correlation between twin
kHz QPOs can be described by regarding the set of oscillations of $n_{\rm r}=0$
and $n_{\rm r}=1$ both with $n=1$ as the twin QPOs (Kato 2011b).
In these disks, however, we could not describe the frequency correlation between 
kHz QPOs and HBOs.
This is because in these disks of $\eta_{\rm s}=\infty$ the frequencies of $n=2$ oscillations are 
not as low as those of observed HBOs.
In the present disks which are terminated at certain finite heights, however, the frequencies of 
$n=2$ oscillations become lower than those in disks of $\eta_{\rm s}=\infty$.
Hence, there is a possibility that the frequency correlation among the kHz QPOs and HBOs can be
also described by the present model in a unified frame.
This possibility will be examined in a subsequent paper (Kato 2012).

Finally, excitation of the vertical p-mode oscillations is mentioned.
We think that not only the vertical p-mode oscillations but also many other trapped
oscillations in disks will be excited by stochastic processes of turbulence 
(Goldreich and Keely 1977a, b) as non-radial oscillations in the Sun and stars, 
although there is no quantitative discussion on this possibility yet.
In stochastic excitation of oscillations by turbulence there will be no particular selection 
of oscillation modes excited, unlike the case of $\kappa$-mechanism of stellar oscillations
or unlike the case of resonant excitation of oscillations.
This might be one of reasons why there are variety of QPOs in disks.

\bigskip
The author thanks the referee for careful reading of the original version with helpful comments.

\bigskip\noindent
{\bf Appendix. Derivation of $A$ and $B$ Given by Equations (\ref{equationA}) and (\ref{equationB})}

By performing integration by part, we have
\begin{equation}
    \biggr\langle g_{n,s}^{(0)}(\eta),\ \eta \frac{d}{d\eta}g_{n,s}^{(0)}(\eta)\biggr\rangle 
       = -\frac{1}{2}I_{n,s,0}+\frac{1}{2}I_{n,s,2}+ S,
\label{appendix1}
\end{equation}
where $I_{n,s,0}$ and $I_{n,s,2}$ are given by equations (\ref{Ins0-Ins2}) and $S$ by equation 
(\ref{surface}).    
Next, considering that $g_{n,s}^{(0)}(\eta)$ is governed by equation (\ref{}), we have,
using equation (\ref{appendix1}),
\begin{equation}
    \biggr\langle g_{n,s}^{(0)}(\eta),\ \frac{d^2}{d\eta^2} g_{n,s}^{(0)}(\eta)\biggr\rangle
     =-\biggr(K_{n,s}+\frac{1}{2}\biggr)I_{n,s,0}
      +\frac{1}{2}I_{n,s,2}+S
\label{appendix2}
\end{equation}
Substituting these expressions into equations (\ref{AA}) and (\ref{B}), we have $A$ and $B$ given by equations
(\ref{equationA}) and (\ref{equationB}).

\bigskip
\leftskip=20pt
\parindent=-20pt
\par
{\bf References}
\par
Goldreich, P. \& Keely, D.A. 1977a, ApJ, 211, 934 \par
Goldreich, P. \& Keely, D.A. 1977b, ApJ, 212, 243 \par
Kato, S. 2001, PASJ, 53, 1\par 
Kato, S. 2010, PASJ, 62, 635 \par
Kato, S. 2011a, PASJ, 63, 125 (paper I)\par
Kato, S. 2011b, PASJ, 63, 861 \par
Kato, S. 2012, PASJ, to be published \par
Kato, S., Fukue, J., \& Mineshige, S. 1998, Black-Hole Accretion Disks 
  (Kyoto: Kyoto University Press), chap. 17 \par
Kato, S., Fukue, J., \& Mineshige, S. 2008, Black-Hole Accretion Disks --- Towards a New paradigm --- 
  (Kyoto: Kyoto University Press), chaps. 3, 11 \par
Machida, M., Nakamura, K.E., \& Matsumoto, R., 2004, PASJ, 56, 671 \par
Oda, H., Machida, M., Nakamura, K.E., Matsumoto, R. 2007, PASJ 59, 457 \par 
Oda, H., Machida, M., Nakamura, K.E., Matsumoto, R. 2009, ApJ, 697, 16 \par 
Oda, H., Machida, M., Nakamura, K.E., Matsumoto, R. 2010, ApJ, 712, 639 \par 
Okazaki, A.T., Kato, S., \& Fukue, J. 1987, PASJ, 39, 457 \par
Ortega-Rodrigues, M., Silbergleit, A.S., Wagoner, R. 2008, Geophys. \& Astrophys. Fluid Dynamics,
     102, 75 \par
Silbergleit, A.S., Wagoner, R., \& Ortega-Rodriguez, M. 2001, ApJ, 548, 335\par
Wagoner, R.V. 1999, Phys. Rev. Rep. 311, 259 \par

\leftskip=20pt
\parindent=0pt
Note added on Feb. 3, 2012:

In this paper we have adopted the approximation that the vertical p-mode oscillations are nearly 
vertical and thus the horizontal motions associated with them are small perturbations over the
vertical ones.
This approximation is qualitatively relevant since $\epsilon(r)$ is smaller than unity in 
the propagation region of the oscillations, as is shown in figure 9.
Quantitatively, however, this approximation is not accurate enough.
In the special case where there is no magnetic field, we can calculate frequencies of trapped,
vertical p-mode oscillations without using the approximation.
Thus, in this special case, we can directly compare the frequencies and their parameter dependence obtained 
in this paper with those derived without the approximation.
This will be done in a subsequent paper.

\end{document}